\documentclass[a4paper,12pt]{article}

\usepackage{jcappub} 
\usepackage{hyperref}
\hypersetup{
    bookmarks=true,         
    unicode=false,          
    pdftoolbar=true,        
    pdfmenubar=true,        
    pdffitwindow=false,     
    pdfstartview={FitH},    
    pdftitle={My title},    
    pdfauthor={Author},     
    pdfsubject={Subject},   
    pdfcreator={Creator},   
    pdfproducer={Producer}, 
    pdfkeywords={keyword1} {key2} {key3}, 
    pdfnewwindow=true,      
    colorlinks=true,       
    linkcolor=red,          
    citecolor=cyan,        
    filecolor=magenta,      
    urlcolor=blue,           
    linktocpage=true
}
\usepackage{subfigure}
\usepackage{pstricks}
\usepackage{color}
\usepackage{amsfonts}
\usepackage{mathrsfs}
\usepackage{epsfig}
\usepackage{amsmath,amssymb,amsthm,graphicx,latexsym}
\usepackage{ulem}

\newcommand{\be}{\begin{equation}}
\newcommand{\ee}{\end{equation}}
\newcommand{\bea}{\begin{eqnarray}}
\newcommand{\eea}{\end{eqnarray}}
\newcommand{\bml}{\begin{subequations}}
\newcommand{\eml}{\end{subequations}}
\newcommand{\bfig}{\begin{figure}}
\newcommand{\efig}{\end{figure}}

\newcommand{\slashed}{\hspace{-1.1ex}/}

\title{
   Modulus stabilization in higher curvature dilaton gravity}

\author[a]{ Sayantan Choudhury}
\author[b]{ Joydip Mitra}
\author[c]{ Soumitra SenGupta}

\affiliation[a]{ Physics and Applied Mathematics Unit, Indian Statistical Institute, 203 B.T. Road, Kolkata 700 108, India,}
\affiliation[b]{ Department of Physics, Scottish Church College, 1 \& 3 Urquhart Square
Kolkata - 700 006, India,}
\affiliation[c]{ Department of Theoretical Physics,
Indian Association for the Cultivation of Science,
2A \& 2B Raja S. C. Mullick Road,
Kolkata - 700 032, India.}

\emailAdd{sayanphysicsisi@gmail.com, jmphys@scottishchurch.ac.in, tpssg@iacs.res.in}

\abstract{We propose a framework of modulus stabilization in two brane warped geometry 
scenario in presence of higher curvature gravity and dilaton in bulk space-time.
 In the prescribed setup we study various features of the stabilized potential for the modulus field, generated by a bulk scalar degrees of freedom with quartic
interactions localized on the two 3-branes placed at the orbifold fixed points.
We determine the parameter space for the gravidilaton and Gauss-Bonnet  
couplings required to stabilize the modulus in such higher curvature dilaton gravity setup.
}

\begin{document} 
\maketitle
\flushbottom

\section{Introduction}

The Gauss-Bonnet (GB) dilaton gravity is known to be an active area of research in theoretical physics
 through decades, which
was proposed to include the perturbative effects within effective theory based on the well known Einstein's 
gravity at the two-loop level \cite{Berg:2005ja,Cicoli:2007xp,Bershadsky:1988mf,Kakushadze:1999bb,Green:1999pu,Roiban:2007jf}. For such theories the two-loop effective coupling signifies the strength of
the self-interaction between the spin 2 graviton degrees of 
freedom below the Ultra-Violet (UV) cut-off of the quantum theory of gravity. 
 Usually such corrections originate naturally in string theory where power expansion in terms of inverse of Regge slope (or string tension) yields the
higher curvature corrections to pure Einstein’s gravity. 
Supergravity, as the low energy limit 
\cite{Nilles:1983ge,Lyth:1998xn,VanNieuwenhuizen:1981ae,Mazumdar:2010sa,Freedman:1976xh,Choudhury:2014sxa,Choudhury:2014uxa,Choudhury:2013jya,Choudhury:2013zna,Choudhury:2012ib,Choudhury:2011sq,Choudhury:2011rz}
 of heterotic string
theory \cite{Green1,Green2,Polchinski1,Polchinski2,Evans:1986ada,Robb:1986de,Gross:1985rr,Candelas:1986tz,Cai:1986sa}, yields the Gauss-Bonnet (GB) term along with dilaton coupling at the leading order correction.
 Consequently it became an active area of interest as a modified theory of gravity.
In the context of black hole it has been shown that GB correction suppresses graviton emission which makes
the black hole more stable.The correction to black hole entropy due to GB term has
also been explored.Moreover in search of extra dimensions, GB dilaton term in a warped braneworld
model has been studied in the context of first kaluza-klein graviton decay channel investigated
by ATLAS group in LHC experiments. Thus the Gauss-Bonnet dilaton gravity as a modified gravity theory has
been studied  extensively in different contexts as a first step to include the higher curvature effects over
Einstein’ gravity. 
 
Stability of the modulus in such models is an important issue from phenomenological point of view.
Goldberger and Wise (GW) \cite{Goldberger:1999uk,Goldberger:1999un,Goldberger:1999wh} first explicitly showed that the dynamics of a five dimensional
bulk scalar field in Randall Sundrum (RS) two brane setup can stabilize the size of the fifth (extra) dimension to a permissible value
to solve the gauge hierarchy problem. In this paper we examine such scenario in the context of higher curvature gravity,
where the usual Einstein's gravity is modified by the perturbative GB coupling and dilaton coupling.
 In this theoretical prescription the stabilized effective potential for the bulk modulus is generated by
the presence of a bulk scalar field with quartic self interactions localized in two 3-branes.
 This results in a  modulus potential which after minimization
  yields a compactification scale in terms of the VEV’s of the
scalar fields at the two branes. This concomitantly solves the gauge hierarchy problem without introducing any fine tuning
of the model parameters in the prescribed theoretical setup. Here we extend this study to include
higher curvature-dilaton term in the bulk space-time where we neglect the 
effects of back reaction of the bulk scalar on the geometry as was done in case
 of the original GW mechanism.Some critical studies have been made in this context \cite{DeWolfe:1999cp,Dey:2006px,
Das:2007mr,Das:2008uw,Maity:2006in}. 
Broad aspects of the moduli stabilization mechanism in higher dimensions 
\cite {Chacko:1999eb,Ponton:2001hq,Charmousis:2004zd, Burgess:2007vi},
specifically in the context of cosmological studies \cite {Cline:2000xn,Kanti:1999nz,Binetruy:2000wn,Ashcroft:2004rs}
 from braneworlds i.e. inflation, dark energy and with
 non minimal scalar fields coupled to the gravity sector have been reported in 
\cite{Ghoroku:2001pi, Lewandowski:2001qp,Lesgourgues:2003mi,Kobayashi:2004aj,Eto:2004yk,Brevik:2004rt,Nunes:2005up,Ichinose:2004ys,Brummer:2005sh}.

The plan of this paper is as follows: In section \ref{l1} we study the framework of the modulus stabilization mechanism
in the context of GB dilaton gravity. First we propose the background model in higher curvature gravity from which we  compute
 the
the expression of the warp factor. Further using this warped solution we  determine the analytical expression for the 
stabilized potential for the bulk modulus field. To check the consistency of our 
present analysis we 
then study our setup in three distinct limiting situations namely in RS limit and limit when
 either of GB coupling or dilaton coupling is present. 
\section{Modulus Stabilization Mechanism in Gauss-Bonnet dilaton gravity}
\label{l1}
Here we generalize the analysis of modulus stabilization mechanism in warped geometry
 in presence of Gauss-Bonnet coupling and gravidilaton coupling in a 5D bulk.
 The background warped geometry model
is proposed by making use of the following sets of assumptions:
\begin{itemize}
 \item The leading order Einstein's gravity sector is modified by the  Gauss-Bonnet 
\cite{Choudhury:2012yh,Choudhury:2012kw,Choudhury:2013qza,Choudhury:2013eoa,Choudhury:2013dia,Choudhury:2013yg,Kim:1999dq,Lee:2000vf,Kim:2000pz,Kim:2000ym}
 and dilaton coupling \cite{Choudhury:2013eoa,Choudhury:2013dia,Choudhury:2013yg,Choudhury:2013aqa}
 which originates from heterotic string theory.
\item The background warped metric has a RS like structure \cite{Randall:1999ee,Randall:1999vf} on a slice of ${\bf AdS_{5}}$ geometry.
      For example,
     from  10-dimensional string model compactified on ${\bf AdS_5\times S^5}$, one typically obtains
     moduli from ${\bf S^5}$ as scalar degrees of freedom. Such moduli can be stabilized by fluxes.
     In our model, which is similar to a 5-dimensional Randall-Sundrum (RS) model, it is assumed that these 
     degrees of freedom are frozen to their VEV and are non-dynamical at the energy scale under consideration \cite{Reece:2010xj}.
     We therefore focus into the slice of ${\bf AdS_5}$ as is done for the 5-dimensional RS model.
\item The dilaton degrees of freedom is assumed to be confined within the bulk.
\item We allow the interaction between dilaton and the 5D bulk cosmological constant via dilaton coupling.
\item The Higgs field is localized at the visible (TeV) brane and the hierarchy problem is resolved via Planck to TeV scale warping.
\item Additionally while determining the values of the model parameters we require
     that the bulk curvature to be less than the five dimensional Planck scale $M_{5}$ so that
     the classical solution of the 5-dimensional gravitational equations can be trusted \cite{Davoudiasl:1999jd,Das:2013lqa}.
\end{itemize}
\subsection{The background setup}
\label{l1a}
We start our discussion with the following 5D action of the two brane warped geometry model \cite{Choudhury:2013yg}:

\be\begin{array}{llll}\label{eq1}
 \displaystyle S=\int d^{5}x \left[\sqrt{-g_{(5)}}\left\{\frac{M^{3}_{(5)}}{2}R_{(5)}+\frac{\alpha_{(5)}M_{(5)}}{2}
\left[R^{ABCD(5)}R^{(5)}_{ABCD}-4R^{AB(5)}R^{(5)}_{AB}+R^{2}_{(5)}\right]\right.\right.\\ \left.\left.
\displaystyle ~~~~~~~~~~~~~~~~~~~~~~~~~~~+\frac{g^{AB}}{2}\partial_{A}\Phi\partial_{B}\Phi-\frac{m^{2}_{\Phi}}{2}\Phi^{2}
+\frac{g^{AB}}{2}\partial_{A}\chi(y)\partial_{B}\chi(y)-2\Lambda_{5}e^{\chi(y)}\right\}
\right.\\ \left. \displaystyle~~~~~~~~~~~~~~~~~~~~~~~~~~~~~~~~~~~~~~~~~~~~~~~~-
\displaystyle \sum^{2}_{i=1}\sqrt{-g^{(i)}_{(5)}}\left[\lambda_{i}\left(\Phi^{2}-{\cal V}^{2}_{i}\right)^{2}+T_{i}\right]\delta(y-y_{i})\right]
\end{array}\ee

with $A,B,C,D=0,1,2,3,4$.
 Here $i$ signifies the brane index, $i=1(\text{hidden})$, $2(\text{visible})$. 
${\cal V}_{i}$ and $\lambda_{i}$ signifies the VEV and self coupling of the bulk scalar fields on the {\it ith} brane where 
$T_{i}$ is the brane tension  and 
 $\Phi$ represent the bulk scalar degrees of freedom. Additionally $\alpha_{5}$ and $\chi(y)$ represent the 
GB coupling and dilaton field.
The background metric describing slice of the ${\bf AdS_{5}}$ is given by,
\be\begin{array}{llllll}\label{eq2}
   \displaystyle ds^{2}_{5}=g_{AB}dx^{A}dx^{B}=e^{-2A(y)}\eta_{\alpha\beta}dx^{\alpha}dx^{\beta}+r^{2}_{c}dy^{2}
   \end{array}\ee
where $r_{c}$ represents the compactification radius of extra dimension.
Here the orbifold points are $y_{i}=[0,\pi]$
and periodic boundary condition is imposed in the closed interval $-\pi\leq y\leq\pi$. After orbifolding, the size of the extra dimensional
interval is $\pi r_{c}$. Moreover in the above metric ansatz
$e^{-2A(y)}$ represents the warp factor while  $\eta_{\alpha\beta}=(-1,+1,+1,+1)$ is flat Minkowski metric.
A more general brane metric for a purely Einsteinian bulk has been discussed in \cite{Koley:2010za}.
\subsection{Warp factor}
\label{l1b}
After varying the model action stated in equation(\ref{eq1}) with respect to the background metric 
the 5D bulk equation of motion turns out to be,
\be\begin{array}{lllll}\label{eq3}
    \displaystyle \sqrt{-g_{(5)}}\left[G^{(5)}_{AB}+\frac{\alpha_{(5)}}{M^{2}_{(5)}}H^{(5)}_{AB}\right]
=-\frac{e^{\chi(y)}}{M^{3}_{(5)}}\left[\Lambda_{(5)} \sqrt{-g_{(5)}}g^{(5)}_{AB}+
\sum^{2}_{i=1}T_{i}\sqrt{-g^{(i)}_{(5)}}g^{(i)}_{\alpha\beta}\delta^{\alpha}_{A}\delta^{\beta}_{B}\delta(y-y_{i})\right]
   \end{array}\ee
where the five dimensional Einstein's tensor and the Gauss-Bonnet tensor are given by

\be\begin{array}{llll}\label{eq4}
    G^{(5)}_{AB}=\left[R^{(5)}_{AB}-\frac{1}{2}g^{(5)}_{AB}R_{(5)}\right],
   \end{array}\ee
and
\be\begin{array}{llll}\label{eq5}
  H^{(5)}_{AB}=2R^{(5)}_{ACDE}R_{B}^{CDE(5)}-4R_{ACBD}^{(5)}R^{CD(5)}
-4R_{AC}^{(5)}R_{B}^{C(5)}+2R^{(5)}R_{AB}^{(5)}\\ ~~~~~~~~~~~~~~~~~~~~~~~~~~~~~~~~~~~~~~~~~~~-\frac{1}{2}g^{(5)}_{AB}
\left(R^{ABCD(5)}R^{(5)}_{ABCD}-4R^{AB(5)}R^{(5)}_{AB}+R^{2}_{(5)}\right).
   \end{array}\ee
Similarly varying equation(\ref{eq1}) with respect to the dilaton field the gravidilaton equation of motion turns out to be
\be\begin{array}{llll}\label{eq6}
\displaystyle \frac{1}{M^{2}_{(5)}}\sum^{2}_{i=1}T_{i}\sqrt{-g^{(i)}_{(5)}}e^{\chi(y)}\delta(y-y_{i})
=\sqrt{-g_{(5)}}\left\{
\displaystyle 
2\frac{\Lambda_{(5)}}{M^{2}_{(5)}}e^{\chi(y)}+\frac{\Box_{(5)} \chi}{M_{(5)}}\right\}
   \end{array}\ee
where the five dimensional D'Alembertian operator is defined as:
\be\label{eq7} \Box_{(5)}\chi(y)=\frac{1}{\sqrt{-g_{(5)}}}\partial_{A}\left(\sqrt{-g_{(5)}}\partial^{A}\chi(y)\right).\ee
Now using the ${\bf Z_{2}}$ orbifolding, we obtain 
at the leading order of $\alpha_{(5)}$  \cite{Choudhury:2013yg}:
\be\begin{array}{llll}\label{gradilatonic}
 \displaystyle  \chi(y)=\left(c_{1}|y|+c_{2} \right) 
   \end{array}\ee
where $c_{1}$ and $c_{2}$ are arbitrary integration constants in which $c_{1}$ characterizes the strength of the dilaton self interaction within the bulk.
 The corresponding warp factor turns out to be \cite{Choudhury:2013yg}:,
\be\begin{array}{llll}\label{warp}
   \displaystyle A(y):= A_{\pm}(y)=k_{\pm}(y)r_{c}|y|\end{array}\ee
where 
\be\begin{array}{llll}\label{wsol}
\displaystyle k_{\pm}(y)=\sqrt{\frac{3M^{2}_{(5)}}{16\alpha_{(5)}}
\left[1\pm\sqrt{\left(1+\frac{4\alpha_{(5)}\Lambda_{5}e^{\chi(y)}}{9M^{5}_{(5)}}\right)}\right]}.\end{array}\ee

 In the small $\alpha_{(5)}$, $c_{1}$ and $c_{2}$ limit we retrieve the results as in the case of RS model with:
\be \label{rs} 
k_{-}(y)\rightarrow k_{RS}=\sqrt{-\frac{\Lambda_{5}}{24M^{3}_{(5)}}}.\ee 

Here we have discarded 
the +ve branch of solution of $k_{+}$ which diverges in the small $\alpha_{(5)}$ limit, bringing 
in ghost fields \cite{Rizzo:2004rq,Dotti:2007az,Torii:2005xu,Konoplya:2010vz,Kim:2000ym,Nojiri:2010wj}. 
Now expanding Eq~(\ref{wsol}) in the perturbation series order by order around 
$\alpha_{5}\rightarrow 0$, $c_{1}\rightarrow 0$ and  $c_{2}\rightarrow 0$ we can write:

 \be\begin{array}{lll}\label{wsol1}
\displaystyle k_{\bf M}(y):=k_{-}(y)=k_{RS}~e^{\frac{\chi(y)}{2}}\left[1+\frac{4\alpha_{(5)} k^{2}_{RS}}{M^{2}_{(5)}}
+{\cal O}\left(\frac{\alpha^{2}_{(5)}k^{4}_{RS}}{M^{4}_{(5)}}\right)+\cdots\right].
\end{array}\ee

\subsection{Stabilized potential for the modulus field}
\label{l1c}

Here we start with the background model action stated in Eq~(\ref{eq1}). 
After varying the Eq~(\ref{eq1}) with respect to the scalar field $\Phi$ we get the 
following equation of motion:
\be\begin{array}{llll}\label{eom1}
    \displaystyle -\frac{1}{r^{2}_{c}}\partial_{y}\left(e^{-4k_{\bf M}(y)r_{c}|y|}\partial_{y}\Phi\right)+m^{2}_{\Phi}e^{-4k_{\bf M}(y)r_{c}|y|}\Phi
+\frac{4}{r_{c}}\sum^{2}_{i=1}e^{-k_{\bf M}(y)r_{c}|y|}\lambda_{i}\Phi\left(\Phi^{2}-{\cal V}^{2}_{i}\right)\delta(y-y_{i})=0
   \end{array}
\ee
which clearly shows that the equation of motion changes from its RS counterpart 
due  an additional coordinate dependence of the function via the dilaton field $\chi(y)$ in $k_{\bf M}(y)$.
For convenience we introduce a set of parameters as:
\be\begin{array}{llll}\label{parameter}
\displaystyle L=\frac{4 \alpha_{(5)} k^{2}_{RS}}{M^{2}_{(5)}},~~~
\displaystyle G=m_{\Phi}^2 r_{c}^2=M_{1}r_{c}\\
\displaystyle S= 4 k_{RS} c_{1} r_{c},~~~
\displaystyle Q=4 k_{RS} r_{c},\\
\displaystyle Z_{L}=(1+L+{\cal O}(L^2)).
\end{array} 
\ee
 Further using Eq.~(\ref{parameter}) in Eq~(\ref{wsol1}) one can re-express the warp function $k_{\bf M}(y)$ as:
\be\label{modified}
k_{\bf M}(y)=k_{RS}~e^{\frac{c_{1}|y|}{2}} Z_{L}.
\ee

 Now solving the Eq~(\ref{eom1}) we obtain the 
solution for the bulk scalar field as,  
\be\begin{array}{llll}\label{soln}
\Phi(y)= A_{1} H_{-A} (B+ BSy) + B_{1}~ {}_1 F_1[\frac{A}{2}, \frac{1}{2},(BQ+BSy)^2] \\
\end{array}
\ee
where,
\be\begin{array}{llll}\label{cont}
   \displaystyle  A=\frac{G}{Z_{L}S},~~~~
B=\frac{\sqrt{Z_{L}}}{\sqrt{2 S}} .
    \end{array}
\ee
 Here ${}_1 F_1$ represents the hypergeometric function of first kind and $H_{-A}$ represents
the Hermite function. Also $A_{1}$ and $B_{1}$ are the arbitrary integration constants which can be evaluated by using
appropriate boundary conditions at the locations of the branes in the prescribed two brane setup.

Since in the perturbative regime of the warping solution the GB coupling $\alpha_{(5)}$ and dilaton coupling $c_1$ is usually small, hence we
can expand the above solution in a series form  and retain upto second order terms which enables us to
recast the solution for the bulk scalar field stated in Eq~(\ref{soln}) as,
 \be\begin{array}{llll} \label{soln1}
 \displaystyle   \Phi(y)= A_1 \frac{ \left[\left\{-2 B(Q+Sy) Z_{L}
 S \Gamma\left[1+\frac{A}{4}\right]\right\}+ \left\{(B^2 G(Q+Sy)^2
+2Z_{L}S)\Gamma[\frac{1}{2}+\frac{A}{4}]\right\}\right]}{2 Z_{L} S \Gamma[1+\frac{A}{2}]}\\
~~~~~~~~~~~~~~~~~~~~~~~~~~~~~~~~~~~~~~~~~~~~~~~~~~~~~~~~~~~~~\displaystyle+B_{1} \left(1+A B^2 (Q+Sy)^2 \right).
    \end{array}
\ee

The effective potential $V_{\Phi}(r_{c})$ can be obtained by substituting the above Eq (\ref{soln1})
into the scalar field action stated in Eq~(\ref{eq1}) and integrating out the extra dimensional
coordinate within $0\leq y\leq\pi$. This results in an effective potential for the modulus $r_c$ which 
is given in the appendix.
 
\subsection{Some limiting cases of Einstein-GB-dilaton model}
We now discuss various limits that can emerge from our proposed model. 
\label{l1d}

\subsubsection{Randall-Sundrum (RS) limit}
\label{as1}
Before discussing the effects of GB and dilaton term let us quickly recall that in absence of these terms
the action corresponds to the stabilization mechanism proposed by Goldberger and Wise. 
 In this case the modulus potential takes the form \cite{Goldberger:1999uk,Goldberger:1999un,Goldberger:1999wh}:
\be \begin{array}{llll} \label{GWpot}
 \displaystyle    V_{\Phi}({r_c})= k_{RS}\epsilon {\cal V}^2_{h}+4k_{RS}e^{-4 k_{RS}r_c \pi}({\cal V}_{v}-{\cal V}_{h}e^{-\epsilon k_{RS}r_c \pi})^2\left(1+\frac{\epsilon}{4}\right)\\
\displaystyle ~~~~~~~~~~~~~~~~~~~~~~~~~~+ k_{RS} \epsilon {\cal V}_{h}e^{-(4+\epsilon) k_{RS}r_c \pi} (2{\cal V}_{v}-{\cal V}_{h} e^{-\epsilon k_{RS}r_c \pi})
\end{array}
 \ee
where $\epsilon=\frac{m^2_{\Phi}}{4 k^2_{RS}} <<1$ for which the terms of ${\cal O} (\epsilon^2)$
can be neglected.

One therefore obtains the minimum of the potential at:
\be\label{minb}
k_{RS}r_c=\frac{4}{\pi}\frac{k^2_{RS}}{m^2_{\Phi}}\ln \left(\frac{{\cal V}_h}{{\cal V}_v}\right)
\ee
Using Eq~(\ref{minb}) one can solve the hierarchy problem by choosing the ratio of VEVs at
 $\frac{{\cal V}_h}{{\cal V}_v}=1.5$ and $\frac{m_{\Phi}}{k_{RS}}=0.2$. This choice
yields $k_{RS}r_{c} \sim 12$.

\subsubsection{Gauss-Bonnet (GB) gravity limit}
\label{as2}  
 In this case we choose the dilation coupling, $c_1= 0$, but the GB coupling $\alpha_{(5)}\neq 0$.
Substituting this in Eq~(\ref{parameter}) we get, $S=0$.
Here, the warp factor takes the form:
\be
k_{M}(y)\rightarrow k_{L}=k_{RS}Z_{L}=k_{RS}(1+L+{\cal O}(L^2))
\ee
This clearly implies that the warp factor in the RS case gets rescaled by a constant factor $Z_{L}=(1+L+{\cal O}(L^2))$ in pure GB limit.
One can obtain the same result as in the case of RS limit by replacing $k_{RS}$ to $k_{L}$ yielding the 
stabilized potential:
\be \begin{array}{llll} \label{GWpotGB}
 \displaystyle    V_{\Phi}({r_c})= k_{L}\epsilon_{L} {\cal V}^2_{h}+4k_{L}e^{-4 k_{L}r_c \pi}({\cal V}_{v}-{\cal V}_{h}e^{-\epsilon_{L} k_{L}r_c \pi})^2\left(1+\frac{\epsilon_{L}}{4}\right)\\
\displaystyle ~~~~~~~~~~~~~~~~~~~~~~~~~~+ k_{L} \epsilon_{L} {\cal V}_{h}e^{-(4+\epsilon_{L}) k_{L}r_c \pi} (2{\cal V}_{v}-{\cal V}_{h} e^{-\epsilon_{L} k_{L}r_c \pi})
\end{array}
 \ee
where $\epsilon_{L}=\frac{m^2_{\Phi}}{4 k^2_{L}} <<1$ for which the terms of ${\cal O} (\epsilon^2_{L})$
has been neglected.
Consequently the 
minima appears at:
\be \label {GWmin}
k_{L}r_{c}=\frac{4}{\pi}\frac{k^2_{L}}{m^2_{\Phi}}ln \left(\frac{{\cal V}_h}{{\cal V}_v}\right)
\Rightarrow k_{RS}r_{c}=\frac{4}{\pi}\frac{k^2_{RS}}{m^2_{\Phi}}Z_{L}\ln \left(\frac{{\cal V}_h}{{\cal V}_v}\right)
\ee
where ${\cal O}(L^2)<<1$ terms can be neglected in the perturbative regime of the solution.
Since $k_{L}$ depends on both $\alpha_{(5)}$ and $k_{RS}$, we can get a family of solutions in terms of $k_{RS}$ and $\alpha_{5}$
to solve the gauge hierarchy problem in the Einstein Gauss-Bonnet gravity. This we shall discuss in a
more general set up later.

\subsubsection{Dilaton gravity limit}
\label{as3}
In this particular case, the GB coupling $\alpha_{(5)}=0$, but the dilaton coupling $c_1\neq 0$, which results in pure dilaton gravity limit.
Substituting this limit in Eq~(\ref{parameter}) we get, $L=0,Z_{L}=1$.
The warp factor in this case takes the form:
\be\label{modifieddil}
k_{\bf M}(y)\rightarrow k_{D}(y)=k_{RS}~e^{\frac{c_{1}|y|}{2}}.
\ee

The classical differential equation for scalar field
in the bulk turn out to be 
\be \begin{array}{llll} \label{GWeqndil} 
    \displaystyle  -\frac{1}{r^2_{c}} \partial_{y}\left( e^{-4 k_{D}(y)r_c |y|} \partial_{y} \Phi(y)\right)+m^2_{\Phi} e^{-4 k_{D}(y)r_c |y|} \Phi(y)
+\frac{4}{r_{c}}\sum^{2}_{i=1}e^{-k_{D}(y)r_{c}|y|}\lambda_{i}\Phi\left(\Phi^{2}-{\cal V}^{2}_{i}\right)\delta(y-y_{i})=0
   \end{array}
\ee
 Away from the boundaries at $y=0,\pi$, the general
solution of Eq~(\ref{GWeqndil}) can be written as:
\be\begin{array}{llll} \label{soln1dil}
 \displaystyle   \Phi(y)= A_1 \frac{ \left[\left\{-2\sqrt{2S} \Gamma\left[\frac{2+\frac{G}{2S}}{4}\right]\right\}+ \left\{(\frac{G}{2S}(Q+Sy)^2
+2S)\Gamma[\frac{1}{2}+\frac{G}{4S}]\right\}\right]}{4S \Gamma[\frac{G}{2S}]}\\
~~~~~~~~~~~~~~~~~~~~~~~~~~~~~~~~~~~~~~~~~~~~~~~~~~~~~~~~~~~~\displaystyle+B_{1} \left(1+\frac{G(Q+Sy)^2}{2S^{2}} \right).
    \end{array}
\ee
This results in an effective potential which is explicitly given in the appendix.

\begin{figure*}[t]
\centering
\subfigure[$L=10^{-7},S=0.09,\frac{{\cal V}_h}{{\cal V}_v}=1.25$]{
    \includegraphics[width=7.2cm, height=8.3cm] {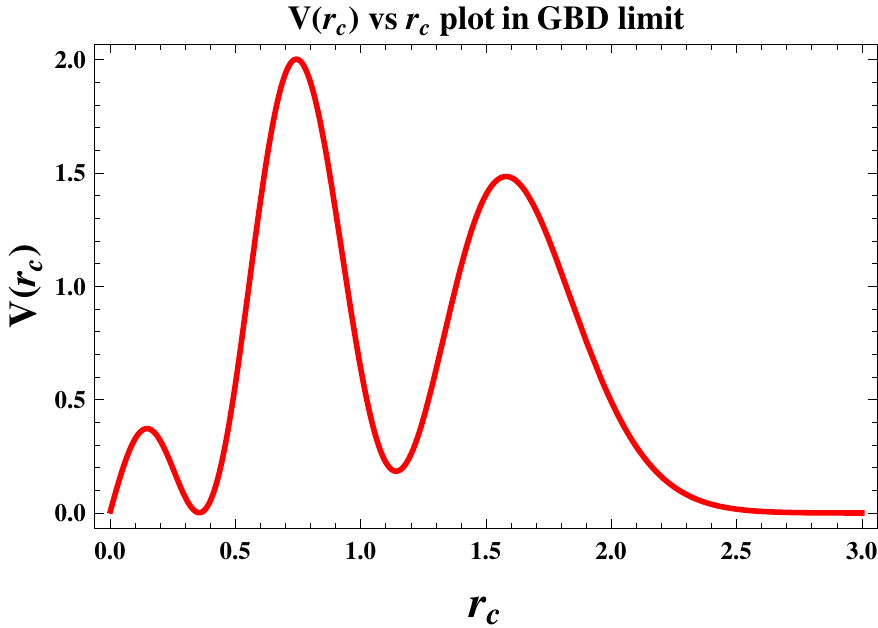}
    \label{fig:subfig1}
}
\subfigure[$L=10^{-1},S=0.09,\frac{{\cal V}_h}{{\cal V}_v}=1.25$]{
    \includegraphics[width=7.2cm, height=8.3cm] {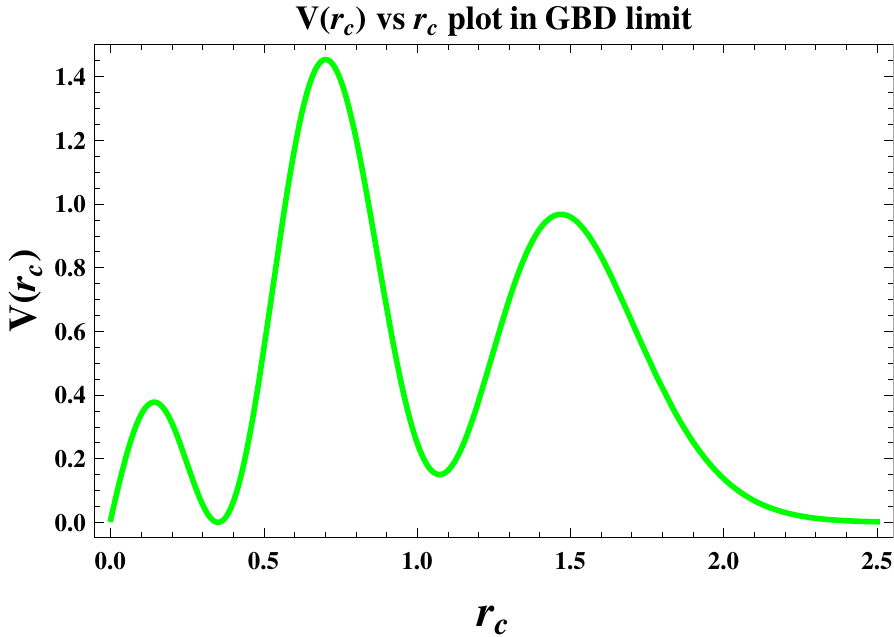}
    \label{fig:subfig2}
}
\subfigure[$L=0.78,S=0.09,\frac{{\cal V}_h}{{\cal V}_v}=1.25$]{
    \includegraphics[width=7.2cm, height=8.3cm] {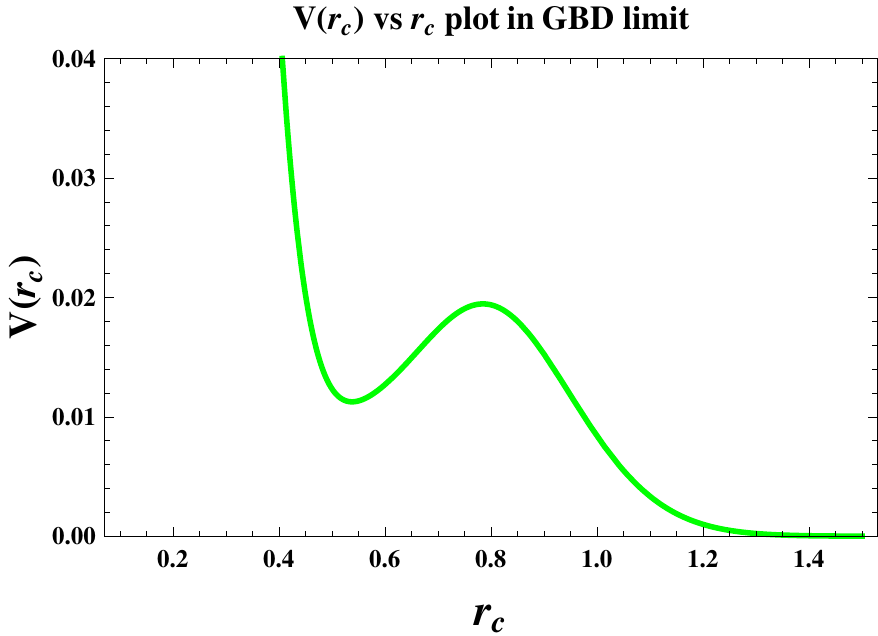}
    \label{fig:subfig3}
}
\subfigure[$L=0.92,S=0.09,\frac{{\cal V}_h}{{\cal V}_v}=1.25$]{
    \includegraphics[width=7.2cm, height=8.3cm] {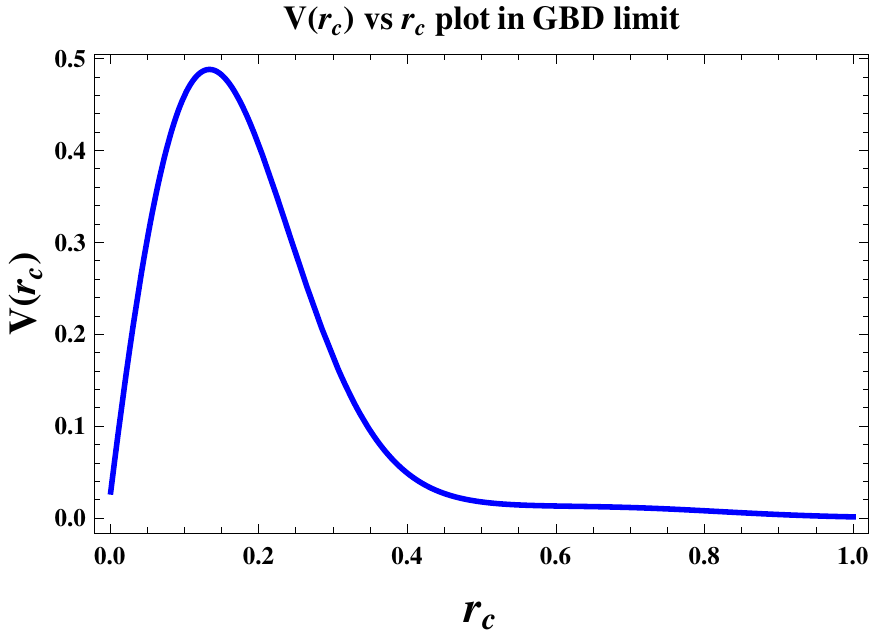}
    \label{fig:subfig4}
}
\caption[Optional caption for list of figures]{Behaviour of the moduli stabilized potential with respect to the compactification radius $r_{c}$ in Gauss Bonnet dilaton (GBD)
limit. 
}
\label{fig1}
\end{figure*}


\begin{figure*}[htb]
\centering
\subfigure[$L=0,S=50,\frac{{\cal V}_h}{{\cal V}_v}=1.25$]{
    \includegraphics[width=11.1cm, height=8.3cm] {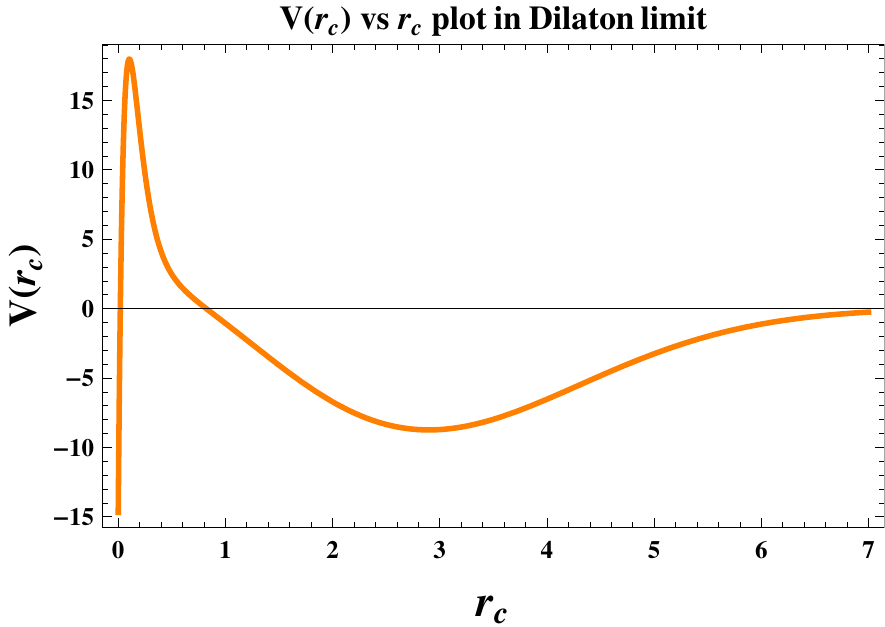}
    \label{fig:subfig6}
}
\subfigure[$L=0,S=0.4,\frac{{\cal V}_h}{{\cal V}_v}=1.25$]{
    \includegraphics[width=11.1cm, height=8.3cm] {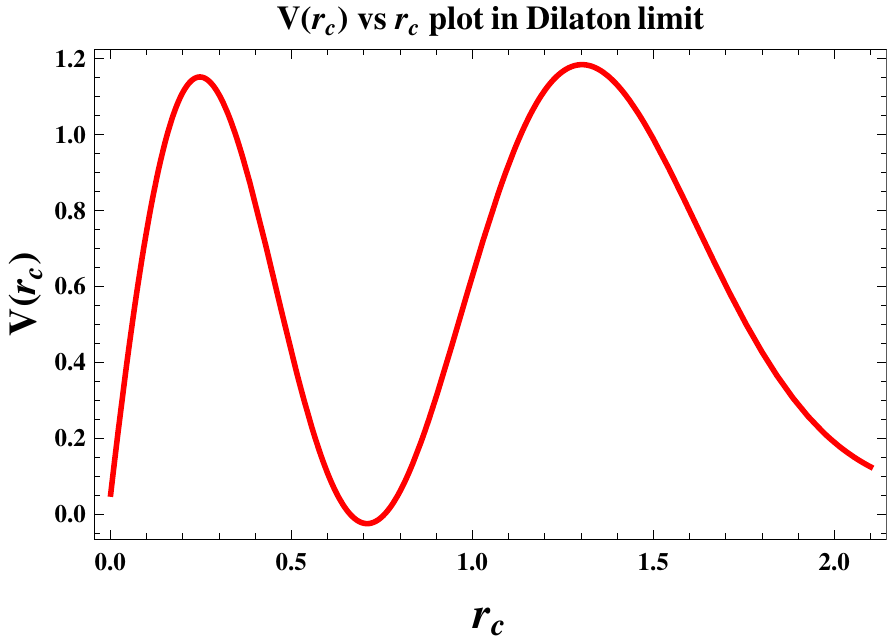}
    \label{fig:subfig7}
}
\caption[Optional caption for list of figures]{Behaviour of the modulus potential with respect to the compactification radius $r_{c}$ in pure dilaton 
limit. 
}
\label{fig2}
\end{figure*}


\begin{figure*}[htb]
\centering
\subfigure[$L=4\times 10^{-7},S=0,\frac{{\cal V}_h}{{\cal V}_v}=1.5$]{
    \includegraphics[width=7.2cm, height=8.3cm] {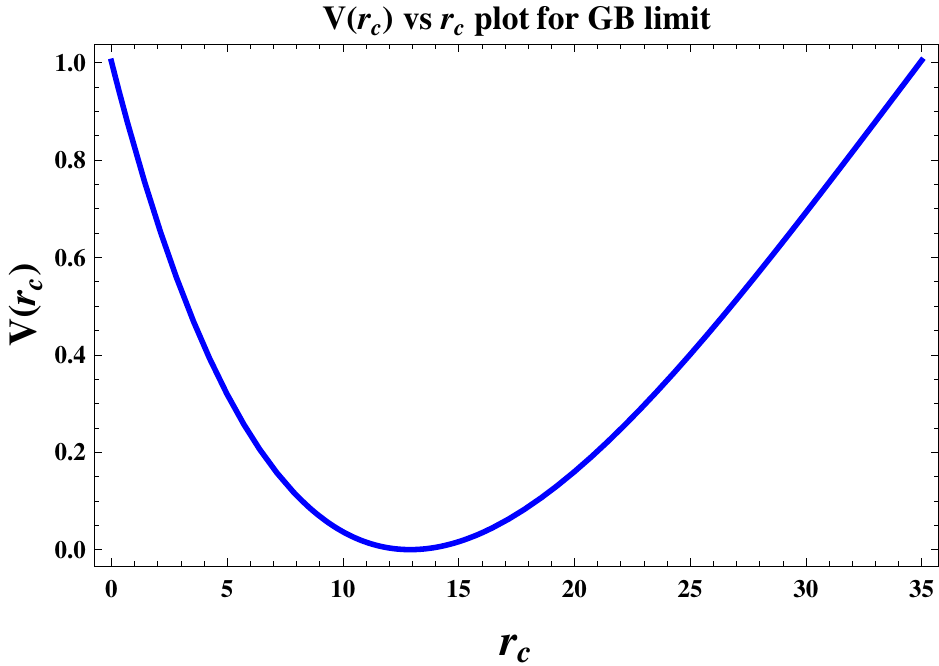}
    \label{fig:subfig8}
}
\subfigure[$L=10^{-1},S=0,\frac{{\cal V}_h}{{\cal V}_v}=1.25$]{
    \includegraphics[width=7.2cm, height=8.3cm] {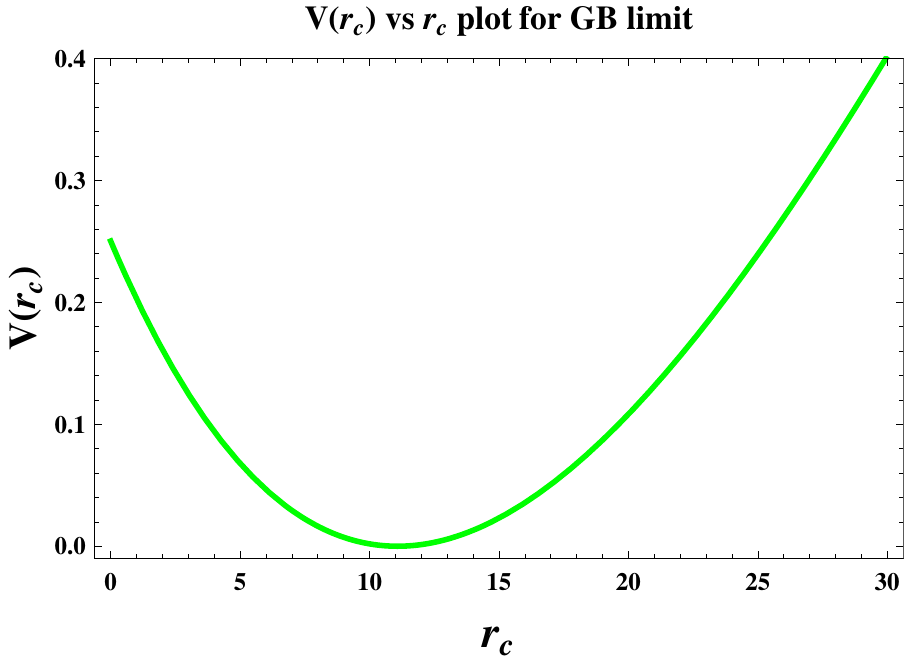}
    \label{fig:subfig8a}
}
\subfigure[$L=0,S=0,\frac{{\cal V}_h}{{\cal V}_v}=1.5$]{
    \includegraphics[width=10.1cm, height=8.3cm] {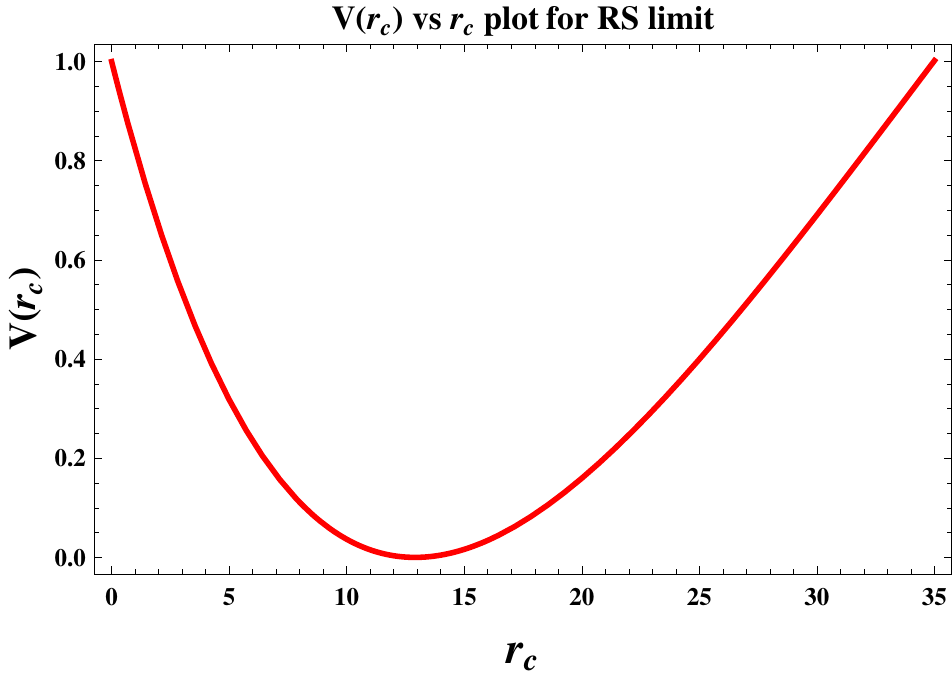}
    \label{fig:subfig9}
}
\caption[Optional caption for list of figures]{Behaviour of the modulus potential with respect to the compactification radius $r_{c}$ 
for Einstein-GB-dilaton gravity.}
\label{fig3}
\end{figure*}

\section{Features of the stabilized potential in higher curvature gravity}

\begin{table}[h]
\renewcommand{\tabcolsep}{0.06pc} 
\small
\begin{tabular}{|c|c|c|c|c|c|c|c|c|c|} \hline
\hline\hline
Different  & L & S  & $\frac{{\cal V}_h}{{\cal V}_v}$& \multicolumn{2}{c|}{existence of } & \multicolumn{2}{c|}{value of $r_c$} & \multicolumn{2}{c|}{value of Potential$V(r_c) $} \\ 
\cline{5-10}
features &  & &  & $minima$ &  $Maxima$ &  $minima$  &  $maxima$  &  $minima$  &  $maxima$   \\ \hline\hline\hline
   Gauss-Bonnet  & $10^{-7}$ & 0.09 & 1.25 & double  &double& 0.3465,1.14 &0.7379,1.573& 0.004842,0.1855& 2.013,1.491 \\ 
Dilaton & $10^{-1}$ & 0.09 & 1.25 & double  & double & 0.3461,1.07 &0.7031, 1.495 & 0.003442,0.1441 & 7.421, 3.621 \\ 
  (GBD limit) & $0.78$ & 0.09 & 1.25 & Single  &X& 0.4975 &X & 0.01214 & X \\ 
  & $0.92$ & 0.09 & 1.25 & X  &Single & X & 0.1281 & X &  0.4865 \\ 
 \hline
Dilaton limit& 0& 50 & 1.25 &Single & single &2.873& 0.1019 & -8.719 & 17.79     \\ 
   & 0&0.4 & 1.25 &Single &double &0.7119&0.2496, 1.312 & -0.01827 & 1.156, 1.192    
\\ \hline
GB limit&$4\times 10^{-7}$ &0 & 1.5 & minima & X &12.77&X & -0.002217& X      \\ 
&$10^{-1}$ &0 & 1.25 & minima & X &11.19&X & -0.001096& X      \\\hline
RS limit   & 0 & 0 & 1.5 &minima & X &12.74&X & -0.00067& X    
 \\ \hline\hline\hline
\end{tabular}
\caption{ Values of moduli radius and moduli potential in GBD, dilaton limit, GB and RS limit. } \label{table1}
\end{table}

\subsection{Case I: Einstein-Gauss Bonnet-dilaton bulk $(\alpha_{5} \neq 0, c_1 \neq 0)$:}
It is clear from the table~(\ref{table1}), fig~(\ref{fig:subfig1}) and fig~(\ref{fig:subfig2}) that in this case there
 exists  multiple (double) number of minima of the modulus potential obtained from the stabilization condition
 of modulus within the interval, $10^{-1}\leq L\leq 10^{-7}$ for fixed 
dilaton coupling at $S\sim 0.09$. In fig~(\ref{fig:subfig1}) and fig~(\ref{fig:subfig2}), the first minima 
appears to be more stable than the second one. The presence of more than one minimum implies the possibility of tunneling
     from one minimum to a more stable one i.e.the one with a lesser value of the moduli potential $V(r_{c})$. From 
     the table~(\ref{table1}) it may be seen that this causes decrease in the value of rc. 
     For a given ${\cal V}_h/{\cal V}_v$, this will result into an increase in the value of the warp factor
     causing an enhancement of the value of the graviton Kaluza Klien (KK) mode masses and decrease
     in the value of the KK graviton coupling to brane fields. As a result the cross section
     for the KK graviton exchange will fall.Though the presence of two minima may imply
   the possibility of tunneling, however as the two minima are separated by a width ${\cal O}(M_{p})$ 
one can rule out the possibility  of tunneling from one stabilized minimum to the adjacent one.
 We  also observe from our analysis that if one increases the ratio of VEV,
then the position of the minimum of the potential slightly shifts toward the higher value of the $r_c$. We have seen that as the strength of the 
GB coupling increases, one passes 
from double minima to single minimum. Most significantly, the increase in GB coupling causes the 
minima to disappear  
while a maximum appears in the moduli potential.This signals disappearance of any stable value for the
modulus implying that large GB coupling leads to instability. See fig~(\ref{fig:subfig4}) for details. Moreover it can be seen that as the VEV decreases (ratio becomes $\sim
1.25$), the potential becomes deeper implying greater stability.  Additionally, for $L=0.78,S=0.09$ and $L=0.92, S=0.09$ we get one minimum and one maximum respectively 
as shown in fig~(\ref{fig:subfig3}-\ref{fig:subfig4}). Also we observe that when $L$ changes from $10^{-7}$
to $10^{-1}$, for $S \sim 0.2-0.9$ we get double minima of the potential. As $S$ increases from 0.9 the
double minima disappears and we have single minimum. On the other hand,
if $S$ decreases from 0.2, at about $S \sim 0.014$, we have an appearance of single minimum in the modulus
potential.We always keep $L$ from $10^{-7}$ 
to $10^{-1}$ since      
$L\geq 1$ is not a feasible value as the perturbative setup will no longer be valid and 
the theory goes to the non-perturbative regime of the solution which is beyond the scope of the present analysis. 

\subsection{Case II: Dilaton limit $(\alpha_{5} = 0, c_1 \neq 0)$:}

If one considers the dilaton limit, then from the table~(\ref{table1}), one single minimum is observed. 
In fig~(\ref{fig:subfig6}) and fig~(\ref{fig:subfig7}) we have depicted such features of stabilized potential with respect to modulus
for the weak and strong dilaton coupling fixed at $S=0.4$ and $S=50$ respectively.
We also observe from the present analysis that as in case of GBD scenario no such double minima appears in the scenario where only dilaton coupling is present.
Moreover as the strength of the dilaton coupling increases, 
stability of the effective potential decreases. 

\subsection{Case III: Gauss-Bonnet limit $(\alpha_{5} \neq 0, c_1 = 0)$:}
In GB limit, only single minimum is observed as mentioned in table~(\ref{table1}). 
The behaviour of the modulus potential is depicted in fig~(\ref{fig:subfig8}) for the ratio of the VEV$\sim 1.5$.
Here we choose the value of the GB coupling $\sim {\cal O}(10^{-7})$ as constrained by various collider (i.e. higgs mass, $H\rightarrow 
\gamma\gamma,\tau\bar{\tau}$ decay \cite{Choudhury:2013eoa} obtained from ATLAS \cite{ATLAS:2013mma,Aad:2012tfa,ATLAS:2011ab} 
and CMS data \cite{Chatrchyan:2013lba}) and solar system observations \cite{Chakraborty:2012sd}.
There is no known dynamical origin of the small value of the Gauss-Bonnet coupling ${\cal O}(10^{−7})$. The consistency of the experimental
     results points towards this value.
We have analyzed that as the VEV decreases (ratio becomes $\sim
1.25$) for a fixed GB coupling, the position of the minimum gets closer to the origin. 
By adjusting the GB parameter $L$ , we can address the well known hierarchy problem. For example,
initially the ratio of VEV is fixed at $1.5$. In such a case $k_{L}r_{c}\sim {\cal O}(12.77)$ 
through which one can solve the hierarchy problem even in the weak 
GB coupling $\sim {\cal O}(10^{-7})$. Now if the ratio of the VEV is decreased to $1.25$ then we observe that $k_{L}r_{c}\sim {\cal O}(6.98)$, which implies that fine tuning problem
cannot be addressed with a very weak GB coupling $\sim {\cal O}(10^{-7})$. But if we increase the GB coupling 
to $\sim {\cal O}(10^{-1})$ within 
the perturbative regime
then even with the decreased value of the ratio of VEV to $1.25$ the gauge hierarchy problem can be
 addressed. See fig~(\ref{fig:subfig8a}) for the details.
 Using the Eq~(\ref{GWmin}) we find that the ratio of the VEV can be expressed in the GB limit as:
 \be \frac{{\cal V}_h}{{\cal V}_v}=e^{\frac{\pi r_c m_{\phi}^2}{4 k_{RS}(1+L)}}.\ee


\begin{figure}[t]
\centering
\includegraphics[width=12.3cm,height=10cm]{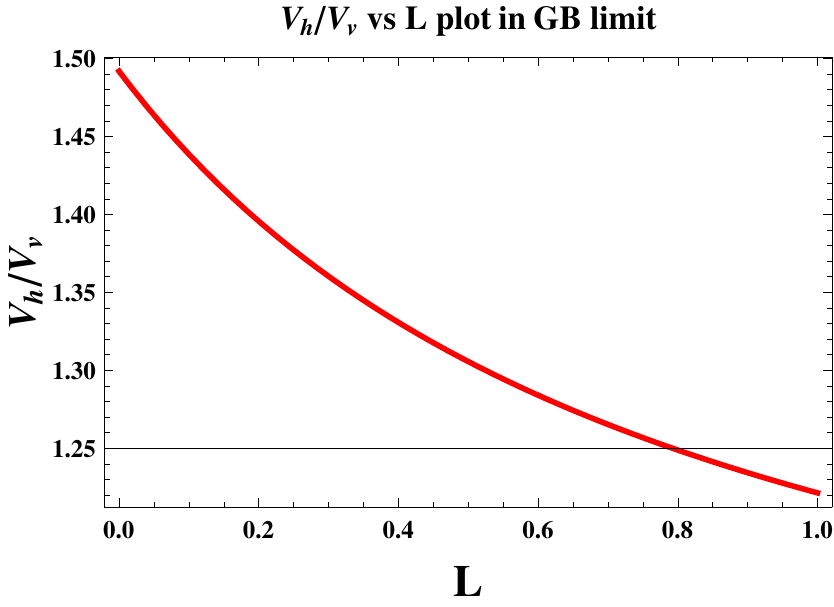}
\caption{\footnotesize Variation of the ratio of VEV with respect to the GB parameter $L$ for $\frac{\pi r_c m_{\phi}^2}{4 k_{RS}}\sim \frac{2}{5}$.
}
\label{figVEV}
\end{figure}

The variation of the ratio of VEV is given with respect to the GB parameter $L$ in Fig~(\ref{figVEV}).From
this figure, it can be clearly seen that in the limit $L \rightarrow 0$,we retrieve the RS limit.
Thus we can generate a parameter space consisting of the GB coupling and ratio of the VEV to 
resolve the hierarchy problem.

Recently, in the context of radion phenomenology,\cite{Maitra:2013cta} it has been shown that in the presence
of GB coupling, radion VEV can be consistently adjusted to give first graviton excitation mass well above 
$\sim$ $3$ TeV as required from the latest ATLAS data.

\subsection{Case IV: Randall-Sundrum limit $(\alpha_{5} = 0, c_1 = 0)$:}
In the RS limit, single minimum has been observed as mentioned in table~(\ref{table1}) The behaviour of the moduli potential is depicted in fig~(\ref{fig:subfig9})
for the ratio of the VEV $\sim 1.5$. To resolve the hierarchy problem, one should fix the 
ratio  to this prescribed value. If the ratio of the VEV decreases( ratio becomes $\sim
1.25$) , the position of the minimum  gets closer to the origin and stability of the effective moduli potential
increases. However unlike the previous case now we have no parameter like GB parameter to the value of 
$k_{RS}r_c$ so that a Planck to TeV scale warping can be achieved. Hence, we can conclude that in case of 
zero GB coupling and zero dialton coupling, we have a specific choice for the ratio of the VEVs of the 
bulk scalar to address the hierarchy issue. The presence of GB and dilaton in the bulk provide us with 
flexibility in this choice.

\section{Conclusion}

In this work, we have studied the modulus stabilization mechanism in warped braneworld model when higher curvature gravity is present in the bulk via GB 
and dilaton coupling (GBD). We have
also studied different limiting situations such as pure GB limit, pure dilaton limit and the RS limit. Analytical expressions for the stabilized
potentials are derived for different cases. We summarize our results as follows:
\begin{itemize}
 \item We observe the existence of double minima when both GB and dilaton coupling are present. As the strength of the GB coupling increases the unstable minimum of these two double minima disappears,
resulting into a single minimum. If we go on increasing the strength of the GB coupling then it is 
observed that the single minimum disappears and a single maximum in the modulus potential appears.
Thus increasing the GB coupling beyond a value leads to instability. Hence, in the perturbative regime of the solution we can always obtain a stabilized modulus potential although these stabilized values
 of the modulus radius $r_c$ are 
not effective in resolving the gauge hierarchy or fine-tuning problem as $k_{\bf M}r_c <<{\cal O}(12)$. 
We observe that as $S$ goes beyond the value $\sim 0.9$ the minimum of the potential disappears
and we move to the region of instability. On the other hand, if value of the dilaton coupling
decreases from a value $\sim 0.2$ we have appearance of single minimum.

\item The existence of double minima of the moduli potential in higher curvature gravity may have interesting consequences in the context of stability of the model.
As the minima in GBD case are separated by a width ${\cal O}(M_{p})$ one can rule out the possibility 
of tunneling from one stabilized minimum to the adjacent one.

\item In case of pure dilaton limit we observe that as the strength of the dilaton coupling increases the stability of the effective moduli potential increases.
Also we have only one minimum of the potential in this case.

\item In case of pure GB limit also only single minimum is observed. For a fixed weak GB coupling, as the ratio of the VEV decreases,
 position of the minimum gets closer
to the origin. It is also observed that using weak GB coupling and large ratio of VEV one cannot solve the hierarchy 
issue. However in the GB limit we observe that if the value of the GB coupling is increased then by decreasing the ratio of VEV  it is 
still possible to resolve the gauge hierarchy problem.

\item In the RS limit single minimum is observed as found in GW mechanism. One can solve the fine-tuning problem by taking a small value of the ratio of the VEV.

\item It is well known that in RS model the various KK graviton modes are important sources for phenomenological 
signatures. The possible diphoton/dilepton decay channel of such gravitons are being studied by 
ATLAS collaboration in LHC.
The most recent result has set stringent lower bounds on the 1st KK graviton $\sim 3$ TeV \cite{Das:2013lqa}.
With pure Einstein gravity in the bulk it is very difficult to satisfy this bound and it has been demonstrated
that the presence of higher curvature terms along with dilaton can explain the
ATLAS result. In this context the study of stability of our proposed model is of utmost importance. 
Through this work we therefore undertake to present a detailed analysis of stabilizing the higher curvature modified
warped geometry model in presence of dilaton.

\end{itemize}

In summary, if we compare our findings with the original Goldberger-Wise stabilization mechanism
we observe that the presence of Gauss-Bonnet (GB) higher curvature term and dilaton term produces the following modification in the 
modulus stabilization scenario. \\ \\

\begin{itemize}
 
\item If GB coupling $L$ increases beyond a desired value, for a given dilaton coupling $S$, then the minima of the potential disappears.\\

\item The value of the dilaton coupling $S$ should be below a critical value to avoid the appearance of 
double minima which removes the possibility of tunneling.\\

\item The reduction in the stabilized value of the modulus $r_c$ (Please see the Table 1) 
than Goldberger-Wise scenario implies an
 improvement in reducing the hierarchy between $r_c$ and inverse of the 4D Planck scale $M^{-1}_{pl}$.

\end{itemize}

\section*{Acknowledgments}

SC thanks Council of Scientific and
Industrial Research, India for financial support through Senior
Research Fellowship (Grant No. 09/093(0132)/2010). SC and JM thanks Indian Association for the Cultivation of Sience (IACS)
where most of the work has been done.
  
\section{Appendix}


Let us explicitly write down the expression for the stabilized potential for the modulus in case of Gauss-Bonnet dilaton:
\be\begin{array}{lllll}\label{potq}
    V_{\Phi}(r_c)=V_{1}(r_c)+V_{2}(r_c),
   \end{array}\ee
where for Gauss-Bonnet dilaton gravity $V_{1}(r_c)$ and $V_{2}(r_c)$ are given by the following expressions:
\be\small\begin{array}{llll} \label{pot1}
V_{1}(r_c)= -\frac{1}{\Gamma[(1+A/4)^2]}M_{1}(-1/(Z_{L}^5 Q^5)e^{-Z_{L}BQ^2}\\ ~~~~~~~~~~~~~~(-M^2(6A^2+6Z_{L}BQ^2 A^2+2Z^3_{L} B Q^4 A(1+B^2Q^2A/2)+Z^4_{L} Q^4(1+B^2Q^2A/2)^2 \\ 
      ~~~~+ 2Z^2_{L} Q^2 A (1+3 B^2Q^2 A/2))\Gamma(1+A/4)^2
-Z_{L}^2 M^2 Q^2 (2+Z_{L}^2 B^2 Q^4+2 Z_{L} B Q^2)A^2 \Gamma[(1+A/2)/2]^2 \\ 
      ~~~~~~~~~~~~~~~~~~+ Z_{L} M Q A (6 A + 6 Z_{L} B Q^2 A+Z_{L}^3 B Q^4 (1+B^2 Q^2 A)\\ ~~~~~~~~~~~~~~~~~+Z_{L}^2 Q^2 (1+3 B^2 Q^2 A))W \Gamma[(1+A/2)/2]\Gamma[(1+A/2)] \\ 
      ~~~~~~~~~~- (24 A^2+24 Z_{L} B Q^2 A^2 +4 Z_{L}^3 B Q^4 A(1+B^2 Q^2 A)+ Z_{L}^4 Q^4 (1+B^2 Q^2 A)^2\\ ~~~~~~~~~~~~+ 4 Z_{L}^2 Q^2 A (1+3 B^2 Q^2 A))W^2 \Gamma[(1+A/2)^2] \\ 
      ~~~~~~~~+ 2 M \Gamma[(1+A/4)](Z_{L}MQA/2 (3A+3 Z_{L} B Q^2 A+Z_{L}^3 B Q^4(1+ B^2 Q^2 A/2)\\ ~~~~~~~~+Z_{L}^2 Q^2 (1+3 B^2 Q^2 A/2)) \Gamma[(1+A/2)/2] - 
      - (12 A^2+12 Z_{L} B Q^2 A^2 + 3 Z_{L}^2 Q^2 A (1+2 B^2 Q^2 A)\\~~~~~~~+ Z_{L}^3 B Q^4 A (3+ 2 B^2 Q^2 A) + Z_{L}^4 Q^4 (1+ 3 B^2 Q^2 A/2 + B^4 Q^4 A^2/2)) W \Gamma[(1+A/2)])) \\ 
      ~~~~~~~+  1/(Z_{L}^5 Q^5)e^{-Z_{L}BQ(Q+\pi S)}(-M^2(6A^2+6Z_{L}BQ (Q+\pi S) A^2\\~~~~~~~~~+2z^3 B Q^3(Q+\pi S) A(1+B^2(Q+\pi S)^2 A/2)
+z^4 Q^4(1+B^2 (Q+\pi S)^2 A/2)^2 \\ 
      ~~~~~~~~~+ 2z^2 Q^2 A (1+3 B^2(Q+\pi S)^2 A/2))\Gamma(1+A/4)^2-(Z_{L}^2 M^2 Q^2/4) (2+2 Z_{L} B Q(Q+\pi S)\\~~~~~~~~+ Z_{L}^2 B^2 Q^2(Q+\pi S))A^2 \Gamma[(1+A/2)/2]^2 
      + Z_{L} M Q A (6 A + 6 Z_{L} B Q(Q+\pi S) A \\ ~~~~~~~~~~+Z_{L}^3 B Q^3(Q+\pi S) (1+B^2 (Q+\pi S)^2 A/2)\\~~~~~~~~~+Z_{L}^2 Q^2 (1+3 B^2 (Q+\pi S)^2 A/2))W \Gamma[(1+A/2)/2]\Gamma[(1+A/2)] \\
      ~~~~~~~- (24 A^2+24 Z_{L} B Q(Q+\pi S) A^2 +4 Z_{L}^3 B Q^3(Q+\pi S) A(1+B^2 (Q+\pi S)^2 A)\\
~~~~~~~~~~~~+ Z_{L}^4 Q^4 (1+B^2 (Q+\pi S)^2 A)^2 + 4 Z_{L}^2 Q^2 A (1+3 B^2 (Q+\pi S)^2 A))W^2 \Gamma[(1+A/2)^2] \\ 
      ~~~~~~~~~~~~~~~~~~+ 2 M \Gamma[(1+A/4)](Z_{L} M Q A/2 (3A+3 Z_{L} B Q(Q+\pi S) A\\ ~~~~+Z_{L}^3 B Q^3(Q+\pi S)(1+ B^2 (Q+\pi S)^2 A/2)+Z_{L}^2 Q^2 (1+3 B^2 (Q+\pi S)^2 A/2)) \Gamma[(1+A/2)/2] - \\ 
      ~~~~~~~~- (12 A^2+12 Z_{L} B Q(Q+\pi S) A^2 + 3 Z_{L}^2 Q^2 A (1+2 B^2 (Q+\pi S)^2 A)\\~~~~~~~~~~ + Z_{L}^3 B Q^3(Q+\pi S) A (3+ 2 B^2 (Q+\pi S)^2 A) + Z_{L}^4 Q^4 (1+ 3 B^2 
(Q+\pi S)^2 A/2 \\~~~~~~~~~~~~+ B^4 (Q+\pi S)^4 A^2/2)) W \Gamma[(1+A/2)])))
      \end{array}
      \ee
      
\be\small\begin{array}{llll} \label{pot2}
 V_{2}(r_c)= -\frac{1}{\Gamma[(1+A/2)]^2} Z_{L} S A ( e^{-Z_{L} B Q^2} (-\frac{1}{4 S}Q A (-4 M \Gamma[(1+A/4)] + 3 Z_{L} M Q \Gamma[(1+A/2)/2]\\
~~~~~~~~~~~~-8 W \Gamma [(1+A/2)] 
          ( M \Gamma[(1+A/4)] + 2 W \Gamma [(1+A/2)])+B^3 Q^3 A(M \Gamma [(1+A/4)]\\
~~~~~~~~~~~~+2 W \Gamma [(1+A/2)])^2-\frac{1}{Z_{L}^3 Q^3} (-4 M^2 A \Gamma [(1+A/4)]^2  
          - Z_{L}^2 M^2 Q^2 A/2 \Gamma[(1+A/2)/2]^2 \\
~~~~~~~~~~~~~+ Z_{L} M Q (Z_{L}^2 Q^2+4 A)W \Gamma[(1+A/2)/2] \Gamma[1+(A/2)]-16 A W^2 \Gamma[1+(A/2)]^2 \\
    ~~~~~~~~~~~~  + M \Gamma[(1+A/4)](Z_{L} M Q(Z_{L}^2 Q^2+2 A) \Gamma[(1+A/2)/2] -16 A W \Gamma[1+(A/2)]))\\
~~~~~~~~~~~~+ \frac{1}{2 Z_{L} B Q S}(2 M^2 (Z_{L}^2 Q^2+2 A)\Gamma[(1+A/4)]^2 
          + Z_{L}^2 M^2 Q^2 A/2 \Gamma[1+(A/2)]^2 \\ 
~~~~~~~~~~~-4 Z_{L} M Q A W \Gamma[(1+(A/2)/2)] \Gamma[1+(A/2)]+ 4(Z_{L}^2 Q^2 +4 A)W^2 \Gamma[1+(A/2)]^2 \\ 
   ~~~~~~~~~~~~~       + 2 M \Gamma[(1+A/4)] (-Z_{L} M Q A \Gamma[(1+(A/2)/2)] + (3 Z_{L}^2 Q^2 +8 A) W \Gamma[(1+(A/2))])))\\ 
  ~~~~~~~~~~~~        +  e^{-Z_{L} B Q(Q+\pi S)} (\frac{1}{4 Q S}(Q+\pi S)^2 A (-4 M \Gamma[(1+A/4)]\\
~~~~~~~~~~~ + 3 Z_{L} M Q \Gamma[(1+A/2)/2]-8 W \Gamma [(1+A/2)])  
          ( M \Gamma[(1+A/4)] + 2 W \Gamma [(1+A/2)])\\
~~~~~~~~~~~~-B^3 (Q+\pi S)^3 A(M \Gamma [(1+A/4)]+2 W \Gamma [(1+A/2)])^2\\
~~~~~~~~~~~~+\frac{1}{Z_{L}^3 Q^3} (-4 M^2 A \Gamma [(1+A/4)]^2  
          - (Z_{L}^2 M^2 Q^2A/2) \Gamma[(1+A/2)/2]^2 \\~~~~~~~~~~~~+ Z_{L} M Q (Z_{L}^2 Q^2+4 A)W \Gamma[(1+A/2)/2] \Gamma[1+(A/2)]-16 A W^2 \Gamma[1+(A/2)]^2 \\ 
          ~~~~~~~~~~~~+ M \Gamma[(1+A/4)](Z_{L} M Q(Z_{L}^2 Q^2+2 A) \Gamma[(1+A/2)/2] -16 A W \Gamma[1+(A/2)]))\\
~~~~~~~~~~~~~- \frac{1}{2 Z_{L} B Q^2 S}(Q+\pi S)(2 M^2 (Z_{L}^2 Q^2+2 A)\Gamma[(1+A/4)]^2  
          + Z_{L}^2 M^2 Q^2 A/2 \Gamma[(1+(A/2))/2]^2 \\
~~~~~~~~~~~-4 Z_{L} M Q A W \Gamma[(1+(A/2)/2)] \Gamma[1+(A/2)]+ 4(Z_{L}^2 Q^2 +4 A)W^2 \Gamma[1+(A/2)]^2 \\ 
          ~~~~~~~~~~~~~+ 2 M \Gamma[(1+A/4)] (-Z_{L} M Q A \Gamma[(1+(A/2)/2)] + (3 Z_{L}^2 Q^2 +8 A) W \Gamma[(1+(A/2))]))))
          \end{array}
          \ee     
where all the parameters $Q,S,L,Z_{L},A,B$ are deined in Eq~(\ref{parameter}) and Eq~(\ref{cont}).
Further if we substitute $Z_{L}=1$ in Eq~(\ref{potq}), Eq~(\ref{pot1}) and Eq~(\ref{pot2}) then it results in the stabilized potential for
 modulus in case of pure dilaton gravity 
limit as mentioned in \ref{as3}.



\begin{thebibliography}{}

\bibitem{Berg:2005ja}
  M.~Berg, M.~Haack and B.~Kors,
  JHEP {\bf 0511} (2005) 030
  [hep-th/0508043].

\bibitem{Cicoli:2007xp}
  M.~Cicoli, J.~P.~Conlon and F.~Quevedo,
  JHEP {\bf 0801} (2008) 052
  [arXiv:0708.1873 [hep-th]].

\bibitem{Bershadsky:1988mf}
  M.~A.~Bershadsky,
  Mod.\ Phys.\ Lett.\ A {\bf 3} (1988) 91.

\bibitem{Kakushadze:1999bb}
  Z.~Kakushadze and T.~R.~Taylor,
  Nucl.\ Phys.\ B {\bf 562} (1999) 78
  [hep-th/9905137].

\bibitem{Green:1999pu}
  M.~B.~Green, H.~-h.~Kwon and P.~Vanhove,
  Phys.\ Rev.\ D {\bf 61} (2000) 104010
  [hep-th/9910055].

\bibitem{Roiban:2007jf}
  R.~Roiban, A.~Tirziu and A.~A.~Tseytlin,
  JHEP {\bf 0707} (2007) 056
  [arXiv:0704.3638 [hep-th]].



\bibitem{Nilles:1983ge}
  H.~P.~Nilles,
  Phys.\ Rept.\  {\bf 110} (1984) 1.

\bibitem{Lyth:1998xn}
  D.~H.~Lyth and A.~Riotto,
  Phys.\ Rept.\  {\bf 314} (1999) 1
  [hep-ph/9807278].

\bibitem{VanNieuwenhuizen:1981ae}
  P.~Van Nieuwenhuizen,
  Phys.\ Rept.\  {\bf 68} (1981) 189.

\bibitem{Mazumdar:2010sa}
  A.~Mazumdar and J.~Rocher,
  Phys.\ Rept.\  {\bf 497} (2011) 85
  [arXiv:1001.0993 [hep-ph]].

\bibitem{Freedman:1976xh}
  D.~Z.~Freedman, P.~van Nieuwenhuizen and S.~Ferrara,
  Phys.\ Rev.\ D {\bf 13} (1976) 3214.


\bibitem{Choudhury:2014sxa}
  S.~Choudhury, A.~Mazumdar and E.~Pukartas,
  JHEP {\bf 1404} (2014) 077
  [arXiv:1402.1227 [hep-th]].

\bibitem{Choudhury:2014uxa}
  S.~Choudhury,
  JHEP {\bf 04} (2014) 105
  [arXiv:1402.1251 [hep-th]].

\bibitem{Choudhury:2013jya}
  S.~Choudhury, A.~Mazumdar and S.~Pal,
  JCAP {\bf 1307} (2013) 041
  [arXiv:1305.6398 [hep-ph]].

\bibitem{Choudhury:2013zna}
  S.~Choudhury, T.~Chakraborty and S.~Pal,
  Nucl.\ Phys.\ B {\bf 880} (2014) 155
  [arXiv:1305.0981 [hep-th]].

\bibitem{Choudhury:2012ib}
  S.~Choudhury and S.~Pal,
  J.\ Phys.\ Conf.\ Ser.\  {\bf 405} (2012) 012009
  [arXiv:1209.5883 [hep-th]].

\bibitem{Choudhury:2011sq}
  S.~Choudhury and S.~Pal,
  Phys.\ Rev.\ D {\bf 85} (2012) 043529
  [arXiv:1102.4206 [hep-th]].

\bibitem{Choudhury:2011rz}
  S.~Choudhury and S.~Pal,
  Nucl.\ Phys.\ B {\bf 857} (2012) 85
  [arXiv:1108.5676 [hep-ph]].

\bibitem{Green1}
  Superstring Theory. Vol. 1: Introduction - Green, Michael B. et al. Cambridge, Uk: Univ. Pr. ( 1987) 469 P. ( Cambridge Monographs On Mathematical Physics).

\bibitem{Green2} 
 Superstring Theory. Vol. 2: Loop Amplitudes,
 Anomalies And Phenomenology - Green, Michael B. et al. Cambridge, Uk: Univ. Pr. ( 1987) 596 P. ( Cambridge Monographs On Mathematical Physics).

\bibitem{Polchinski1}
String theory. Vol. 1: An introduction to the bosonic string - Polchinski, J. Cambridge, UK: Univ. Pr. (1998) 402 p.

\bibitem{Polchinski2}
String theory. Vol. 2: Superstring theory and beyond - Polchinski, J. Cambridge, UK: Univ. Pr. (1998) 531 p.


\bibitem{Evans:1986ada}
  M.~Evans and B.~A.~Ovrut,
  Phys.\ Lett.\ B {\bf 171} (1986) 177.

\bibitem{Robb:1986de}
  T.~D.~Robb and J.~G.~Taylor,
  Phys.\ Lett.\ B {\bf 176} (1986) 355.

\bibitem{Gross:1985rr}
  D.~J.~Gross, J.~A.~Harvey, E.~J.~Martinec and R.~Rohm,
  Nucl.\ Phys.\ B {\bf 267} (1986) 75.

\bibitem{Candelas:1986tz}
  P.~Candelas, M.~D.~Freeman, C.~N.~Pope, M.~F.~Sohnius and K.~S.~Stelle,
  Phys.\ Lett.\ B {\bf 177} (1986) 341.

\bibitem{Cai:1986sa}
  Y.~Cai and C.~A.~Nunez,
  Nucl.\ Phys.\ B {\bf 287} (1987) 279.

\bibitem{Goldberger:1999uk}
  W.~D.~Goldberger and M.~B.~Wise,
  Phys.\ Rev.\ Lett.\  {\bf 83} (1999) 4922
  [hep-ph/9907447].


\bibitem{Goldberger:1999un}
  W.~D.~Goldberger and M.~B.~Wise,
  Phys.\ Lett.\ B {\bf 475} (2000) 275
  [hep-ph/9911457].

\bibitem{Goldberger:1999wh}
  W.~D.~Goldberger and M.~B.~Wise,
  Phys.\ Rev.\ D {\bf 60} (1999) 107505
  [hep-ph/9907218].

\bibitem{DeWolfe:1999cp}
  O.~DeWolfe, D.~Z.~Freedman, S.~S.~Gubser and A.~Karch,
  Phys.\ Rev.\ D {\bf 62} (2000) 046008
  [hep-th/9909134].

\bibitem{Dey:2006px}
  A.~Dey, D.~Maity and S.~SenGupta,
  Phys.\ Rev.\ D {\bf 75} (2007) 107901
  [hep-th/0611262].

\bibitem{Das:2007mr}
  S.~Das, A.~Dey and S.~SenGupta,
  Europhys.\ Lett.\  {\bf 83} (2008) 51002
  [arXiv:0704.3119 [hep-th]].

\bibitem{Das:2008uw}
  A.~Das, S.~Kar and S.~SenGupta,
  Int.\ J.\ Mod.\ Phys.\ A {\bf 24} (2009)
  [arXiv:0804.1757 [hep-th]].

\bibitem{Maity:2006in}
  D.~Maity, S.~SenGupta and S.~Sur,
  Class.\ Quant.\ Grav.\  {\bf 26} (2009) 055003
  [hep-th/0609171].


\bibitem{Chacko:1999eb}
  Z.~Chacko and A.~E.~Nelson,
  Phys.\ Rev.\ D {\bf 62} (2000) 085006
  [hep-th/9912186]

\bibitem{Ponton:2001hq}
  E.~Ponton and E.~Poppitz,
  JHEP {\bf 0106} (2001) 019
  [hep-ph/0105021].

\bibitem{Charmousis:2004zd}
  C.~Charmousis and U.~Ellwanger,
  JHEP {\bf 0402} (2004) 058
  [hep-th/0402019].

\bibitem{Burgess:2007vi}
  C.~P.~Burgess, D.~Hoover and G.~Tasinato,
  JHEP {\bf 0709} (2007) 124
  [arXiv:0705.3212 [hep-th]].

\bibitem{Cline:2000xn}
  J.~M.~Cline and H.~Firouzjahi,
  Phys.\ Rev.\ D {\bf 64} (2001) 023505
  [hep-ph/0005235].

\bibitem{Kanti:1999nz}
  P.~Kanti, I.~I.~Kogan, K.~A.~Olive and M.~Pospelov,
  Phys.\ Rev.\ D {\bf 61} (2000) 106004
  [hep-ph/9912266].


\bibitem{Binetruy:2000wn}
  P.~Binetruy, J.~M.~Cline and C.~Grojean,
  Phys.\ Lett.\ B {\bf 489} (2000) 403
  [hep-th/0007029].


\bibitem{Ashcroft:2004rs}
  P.~R.~Ashcroft, C.~van de Bruck and A.~-C.~Davis,
  Phys.\ Rev.\ D {\bf 71} (2005) 023508
  [astro-ph/0408448].

\bibitem{Ghoroku:2001pi}
  K.~Ghoroku and A.~Nakamura,
  Phys.\ Rev.\ D {\bf 64} (2001) 084028
  [hep-th/0103071].

\bibitem{Lewandowski:2001qp}
  A.~Lewandowski and R.~Sundrum,
  Phys.\ Rev.\ D {\bf 65} (2002) 044003
  [hep-th/0108025].

\bibitem{Lesgourgues:2003mi}
  J.~Lesgourgues and L.~Sorbo,
  Phys.\ Rev.\ D {\bf 69} (2004) 084010
  [hep-th/0310007].

\bibitem{Kobayashi:2004aj}
  T.~Kobayashi and K.~Yoshioka,
  JHEP {\bf 0411} (2004) 024
  [hep-ph/0409355].
\bibitem{Eto:2004yk}
  M.~Eto, N.~Maru and N.~Sakai,
  Phys.\ Rev.\ D {\bf 70} (2004) 086002
  [hep-th/0403009].
\bibitem{Brevik:2004rt}
  I.~H.~Brevik, K.~Ghoroku and M.~Yahiro,
  Phys.\ Rev.\ D {\bf 70} (2004) 064012
  [hep-th/0402176].
\bibitem{Nunes:2005up}
  N.~J.~Nunes and M.~Peloso,
  Phys.\ Lett.\ B {\bf 623} (2005) 147
  [hep-th/0506039].

\bibitem{Ichinose:2004ys}
  S.~Ichinose and A.~Murayama,
  Phys.\ Lett.\ B {\bf 625} (2005) 106
  [hep-th/0409193].

\bibitem{Brummer:2005sh}
  F.~Brummer, A.~Hebecker and E.~Trincherini,
  Nucl.\ Phys.\ B {\bf 738} (2006) 283
  [hep-th/0510113].



\bibitem{Choudhury:2012yh}
  S.~Choudhury and S.~Pal,
  Nucl.\ Phys.\ B {\bf 874} (2013) 85
  [arXiv:1208.4433 [hep-th]].

\bibitem{Choudhury:2012kw}
  S.~Choudhury and S.~Pal,
  arXiv:1210.4478 [hep-th].

\bibitem{Choudhury:2013qza}
  S.~Choudhury and A.~Dasgupta,
  Nucl.\ Phys.\ B {\bf 882} (2014) 195
  [arXiv:1309.1934 [hep-ph]].

\bibitem{Choudhury:2013eoa}
  S.~Choudhury, S.~Sadhukhan and S.~SenGupta,
  arXiv:1308.1477 [hep-ph].

\bibitem{Choudhury:2013dia}
  S.~Choudhury and S.~SenGupta,
  arXiv:1306.0492 [hep-th].

\bibitem{Choudhury:2013yg}
  S.~Choudhury and S.~SenGupta,
  JHEP {\bf 1302} (2013) 136
  [arXiv:1301.0918 [hep-th]].

\bibitem{Choudhury:2013aqa}
  S.~Choudhury and S.~SenGupta,
  arXiv:1311.0730 [hep-ph].

\bibitem{Kim:1999dq}
  J.~E.~Kim, B.~Kyae and H.~M.~Lee,
  Phys.\ Rev.\ D {\bf 62} (2000) 045013
  [hep-ph/9912344].

\bibitem{Lee:2000vf}
  H.~M.~Lee,
  hep-th/0010193.

\bibitem{Kim:2000pz}
  J.~E.~Kim, B.~Kyae and H.~M.~Lee,
  Nucl.\ Phys.\ B {\bf 582} (2000) 296
   [Erratum-ibid.\ B {\bf 591} (2000) 587]
  [hep-th/0004005].

\bibitem{Kim:2000ym}
  J.~E.~Kim and H.~M.~Lee,
  Nucl.\ Phys.\ B {\bf 602} (2001) 346
   [Erratum-ibid.\ B {\bf 619} (2001) 763]
  [hep-th/0010093].

\bibitem{Randall:1999ee}
  L.~Randall and R.~Sundrum,
  Phys.\ Rev.\ Lett.\  {\bf 83} (1999) 3370
  [hep-ph/9905221].

\bibitem{Randall:1999vf}
  L.~Randall and R.~Sundrum,
  Phys.\ Rev.\ Lett.\  {\bf 83} (1999) 4690
  [hep-th/9906064].

\bibitem{Reece:2010xj}
M.~Reece and L.~-T.~Wang,
  JHEP {\bf 1007} (2010) 040
  [arXiv:1003.5669 [hep-ph]].


\bibitem{Davoudiasl:1999jd}
  H.~Davoudiasl, J.~L.~Hewett and T.~G.~Rizzo,
  Phys.\ Rev.\ Lett.\  {\bf 84} (2000) 2080
  [hep-ph/9909255].

\bibitem{Das:2013lqa}
  A.~Das and S.~SenGupta,
  arXiv:1303.2512 [hep-ph].

\bibitem{Koley:2010za}
  R.~Koley, J.~Mitra and S.~SenGupta,
  Europhys.\ Lett.\  {\bf 91} (2010) 31001
  [arXiv:1001.2666 [hep-th]].

\bibitem{Rizzo:2004rq}
  T.~G.~Rizzo,
  JHEP {\bf 0501} (2005) 028
  [hep-ph/0412087].

\bibitem{Dotti:2007az}
  G.~Dotti, J.~Oliva and R.~Troncoso,
  Phys.\ Rev.\ D {\bf 76} (2007) 064038
  [arXiv:0706.1830 [hep-th]].

\bibitem{Torii:2005xu}
  T.~Torii and H.~Maeda,
  Phys.\ Rev.\ D {\bf 71} (2005) 124002
  [hep-th/0504127].

\bibitem{Konoplya:2010vz}
  R.~A.~Konoplya and A.~Zhidenko,
  Phys.\ Rev.\ D {\bf 82} (2010) 084003
  [arXiv:1004.3772 [hep-th]].

\bibitem{Nojiri:2010wj}
  S.~'i.~Nojiri and S.~D.~Odintsov,
  Phys.\ Rept.\  {\bf 505} (2011) 59
  [arXiv:1011.0544 [gr-qc]].

\bibitem{ATLAS:2013mma}
  [ATLAS Collaboration],
  ATLAS-CONF-2013-014.

\bibitem{Aad:2012tfa}
  G.~Aad {\it et al.}  [ATLAS Collaboration],
  Phys.\ Lett.\ B {\bf 716} (2012) 1
  [arXiv:1207.7214 [hep-ex]].

\bibitem{ATLAS:2011ab}
  G.~Aad {\it et al.}  [ATLAS Collaboration],
  Phys.\ Lett.\ B {\bf 710} (2012) 538
  [arXiv:1112.2194 [hep-ex]].

\bibitem{Chatrchyan:2013lba}
  S.~Chatrchyan {\it et al.}  [CMS Collaboration],
  JHEP {\bf 1306} (2013) 081
  [arXiv:1303.4571 [hep-ex]].



\bibitem{Chakraborty:2012sd}
  S.~Chakraborty and S.~SenGupta,
  Phys.\ Rev.\ D {\bf 89} (2014) 026003
  [arXiv:1208.1433 [gr-qc]].

\bibitem{Maitra:2013cta}
  U.~Maitra, B.~Mukhopadhyaya and S.~SenGupta,
  arXiv:1307.3018.




\end{thebibliography}
\end{document}